\documentclass{appolb}
\usepackage[dvips]{epsfig}
\usepackage[dvips]{graphicx}

\begin{document}
\pagestyle{plain}
\newcount\eLiNe\eLiNe=\inputlineno\advance\eLiNe by -1
\title{Jet Physics at CDF}
\author{Kenichi Hatakeyama\thanks{{\tt hatakek@mail.rockefeller.edu}}%
\\
for the CDF Collaboration
\address{The Rockefeller University,
1230 York Avenue, New York, NY, U.S.A.}}
\maketitle

\begin{abstract}
Recent results on jet physics at the Fermilab Tevatron
$p\bar p$ collider from the CDF Collaboration are presented.
The main focus is put on results for the inclusive jet and dijet,
$b\bar b$ dijet, $W/Z+$jets and $W/Z+b$-jets production.
\end{abstract}

\section{Inclusive and dijet production}

The differential inclusive jet cross section and dijet cross
section at the Tevatron test QCD at the shortest distances currently
attainable in accelerator experiments.
The measurements provide a fundamental test of QCD and a constraint on
the parton distribution functions (PDFs) of the proton.
The dijet mass spectrum is also sensitive to the presence of new
particles that decay into two jets.

\begin{figure}[thb]
\begin{center}
\includegraphics[width=0.495\hsize]{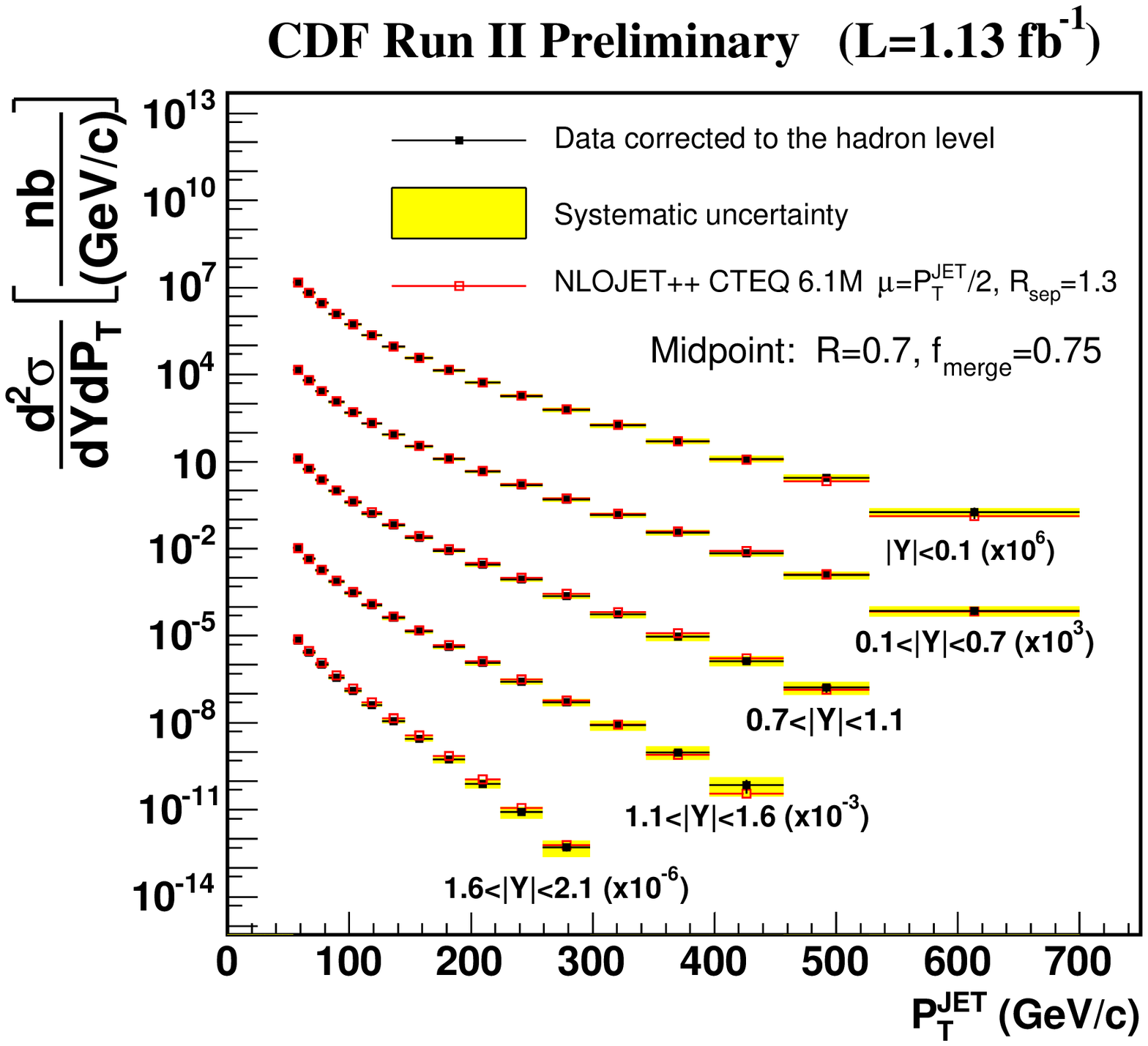}
\includegraphics[width=0.495\hsize,bb=0 -20 567 410]{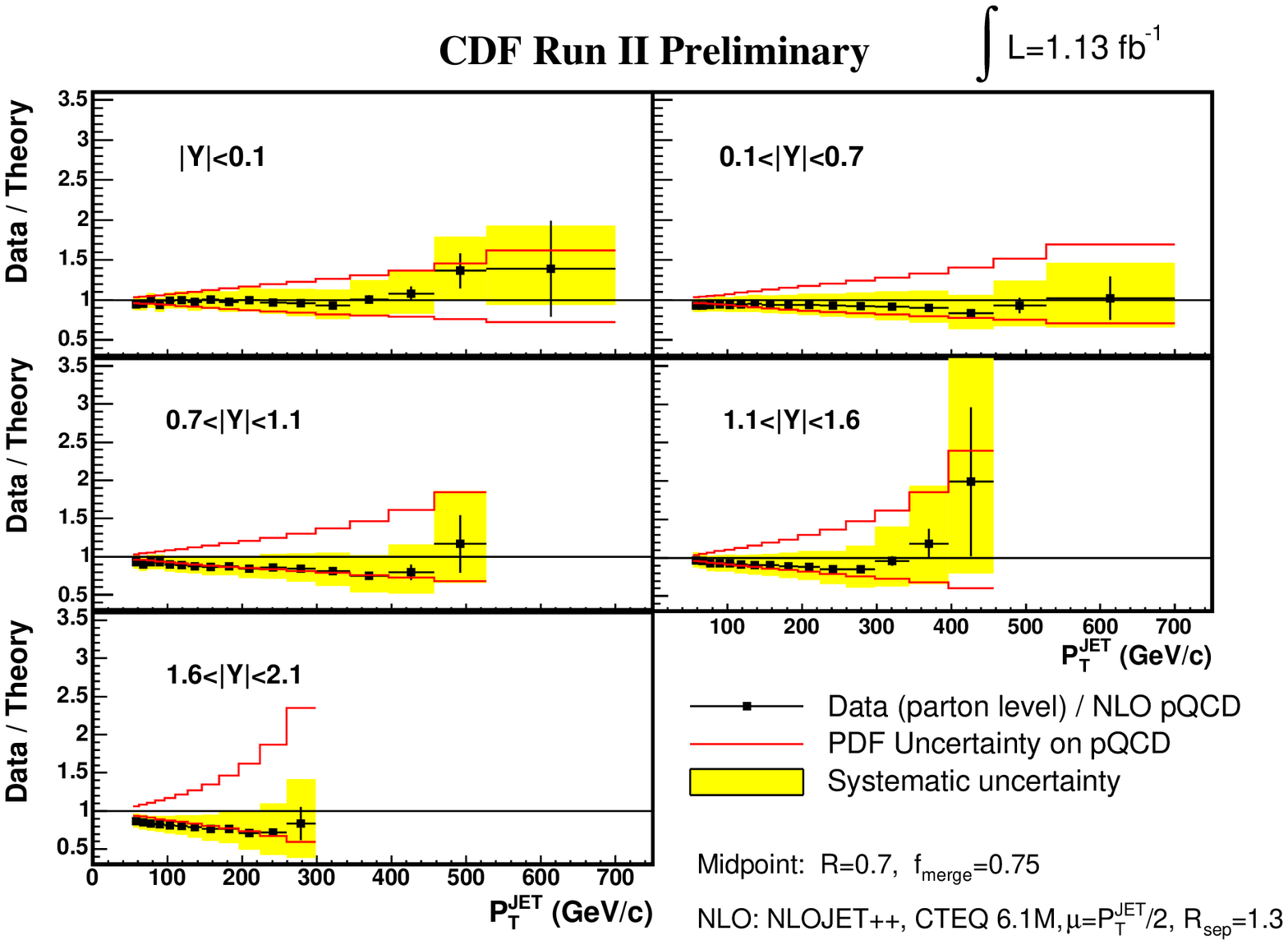}
\caption{{\em (left)} measured inclusive jet differential cross sections
in five rapidity regions compared to NLO pQCD predictions;
{\em (right)} ratios of the measured cross sections
over the NLO pQCD predictions.\label{fig:incjet}}
\end{center}
\end{figure}

CDF has made inclusive jet cross section measurements using the $k_T$
algorithm and Midpoint cone clustering algorithm.
The recent measurements using these algorithms are
based on the 1.0 and 1.13 fb${}^{-1}$ of data, respectively, and cover
the rapidity region up to $|y_{jet}|=2.1$ which is much wider than
$0.1<|y_{jet}|<0.7$ in the previous measurements.
The $k_T$ measurement was published in~\cite{CDFIncJetKt} and
the results from the Midpoint measurement are shown in
Fig.~\ref{fig:incjet}.
The measured cross sections are in agreement with next-to-leading
order (NLO) perturbative QCD (pQCD) predictions based on CTEQ6.1M PDF.
The measurements in the forward region show that the experimental
uncertainties are somewhat smaller than the PDF uncertainties, and
this measurement is expected to further constrain the PDFs.

The dijet mass differential cross section was measured using the
1.13~$\mbox{fb}^{-1}$ of data with the Midpoint algorithm.
The measured dijet mass spectrum was found to be in good agreement
with NLO pQCD predictions within the uncertainties.
Limits on the cross sections for new particles decaying into two jets
have been worked out using this measurement.

\section{$b\bar b$ dijet production}

The $b\bar b$ dijet production cross section has been measured using
260 pb$^{-1}$ of data collected by triggering on two displaced
tracks and two jets. 
$b\bar b$ dijet events are selected by requiring 
two jets with $|\eta_{1,2}|<1.2$ and $E_{T,1}>35$ GeV and $E_{T,2}>32$
GeV, respectively, which are tagged as $b$-jets by a secondary vertex
algorithm.
The measured $b\bar b$ differential cross section is 
compared to Pythia Tune A~\cite{tuneA}, Herwig with
Jimmy~\cite{Jimmy}, and MC@NLO~\cite{MCNLO} with and without Jimmy as
a function of $\Delta\phi$.
Tune A refers to a special set of Pythia parameters tuned to give a
reasonable description of the underlying event (UE), and
Jimmy is a program which can be used in Herwig and MC@NLO to add
multiple parton interactions to events to improve the description of
UE.
MC@NLO+Jimmy provides the best description of the data indicating
the importance of both the next-to-leading order contribution and 
UE effect.

\section{$W/Z$+jets and $W/Z+b$-jets production}

The $W/Z$+jets and $W/Z+b$-jets production has been studied
intensively at CDF.
The studies of these processes provide important tests of pQCD
predictions at high momentum transfers.
Final states containing $W/Z$ and ($b$-)jets are signal
channels for many interesting processes such as $t\bar t$ or single top
production, as well as searches for the Higgs boson in the $W/Z+H\to
W/Z+b\bar b$ channel and physics beyond the Standard Model (SM) such
as Supersymmetry.
The production of $W/Z$+jets via QCD constitutes a large background to
these processes, and thus it is essential to understand these
processes accurately.

The $W$+jets cross sections were measured using $W\to e\nu$
events from the 320 pb${}^{-1}$ of data
for four inclusive jet multiplicities ($N_{jets}\ge1,2,3,4$), and
compared to NLO pQCD predictions from MCFM~\cite{MCFM}
and LO matrix element (ME) + parton shower (PS) Monte Carlo
predictions based on the CKKW~\cite{CKKW} and MLM (as in
Alpgen~\cite{Alpgen}) matching schemes.
The LO ME+PS predictions are systematically lower than the measured
cross sections; however, all the predictions are in agreement with
data in the cross section ratios $\sigma_{n}/\sigma_{n-1}$, where
$\sigma_n=\sigma(W\to e\nu+\ge n\mbox{-jet}; E_{T,\,n{\mbox{\scriptsize
      -th jet}}}\ge25\mbox{ GeV})$.

\begin{figure}[tb]
\begin{center}
\includegraphics[width=0.495\hsize]{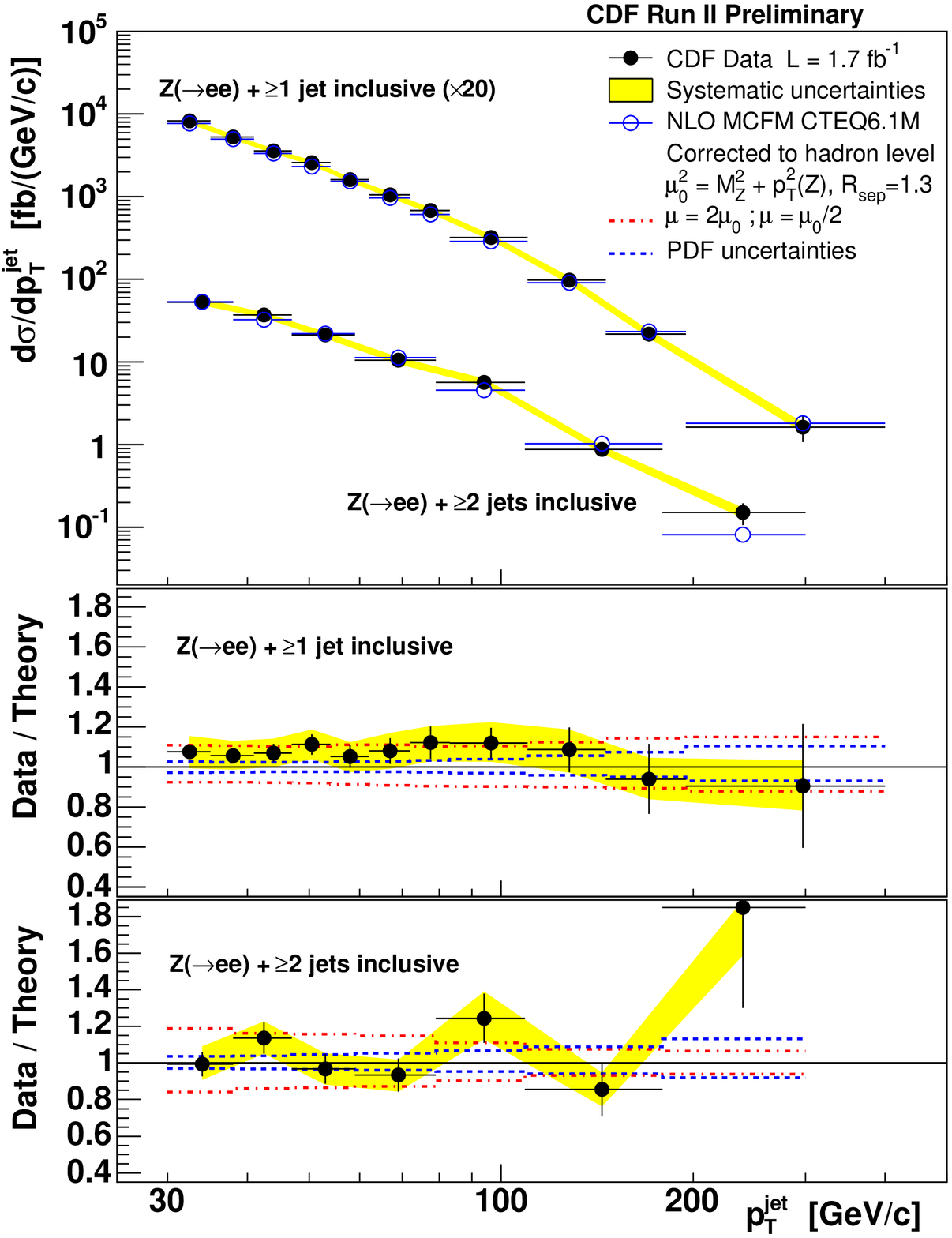}
\includegraphics[width=0.495\hsize,bb=0 -100 567 543]{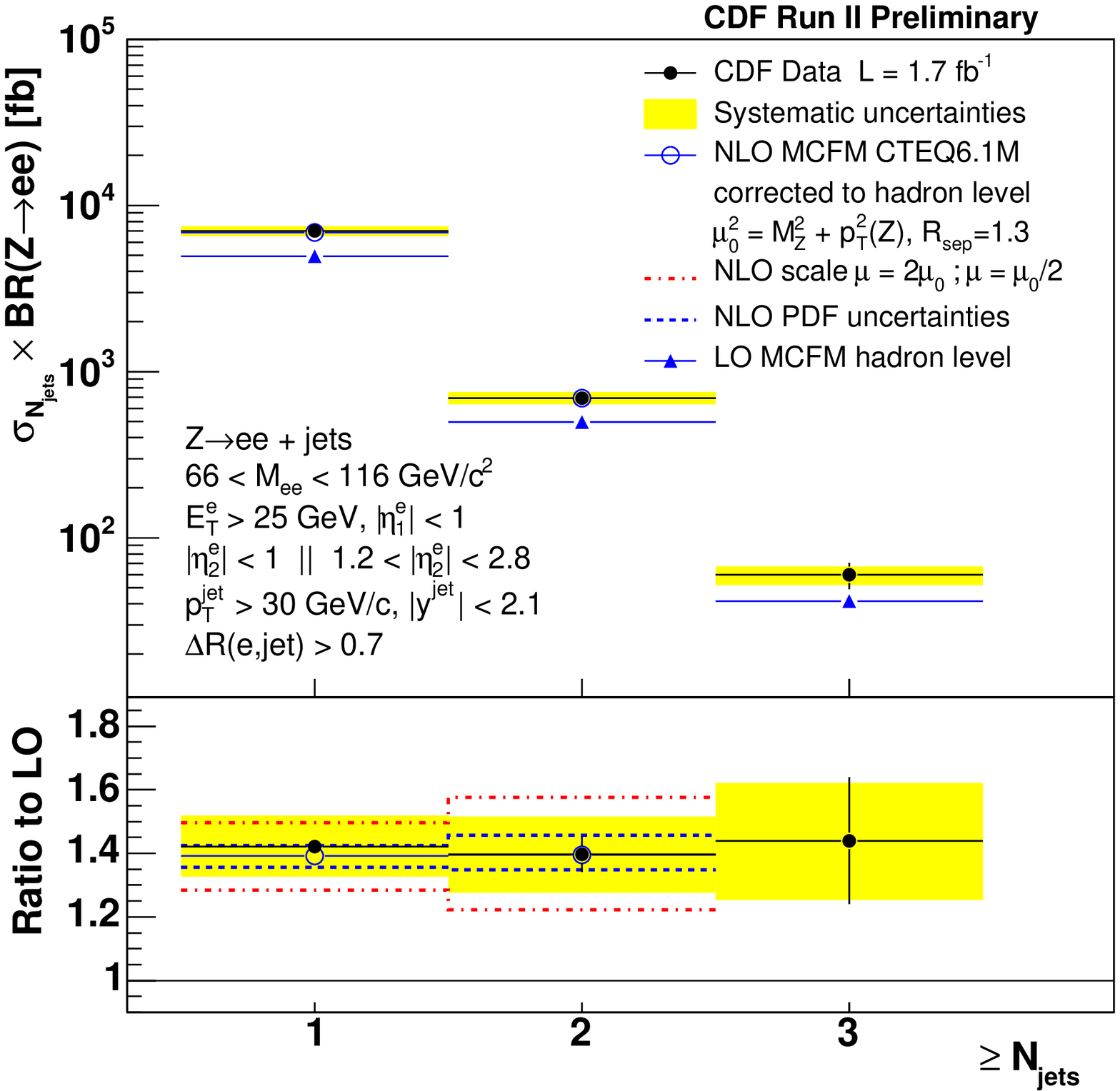}
\caption{{\em (left)}  Measured inclusive cross section for $Z$+jets
  production as a function of $p_T^{jet}$ compared to NLO pQCD
  predictions. 
  {\em (right)} Measured cross section as a function of inclusive jet
  multiplicity compared to NLO pQCD predictions.
  \label{fig:zjets}}
\end{center}
\end{figure}

The $Z$+jets cross sections were measured using $Z\to e^+e^-$
events from the 1.7 fb${}^{-1}$ of data for jets 
in the kinematic region of $p_T^{jet}>30$ GeV/{\it c} and $|y_{jet}|<2.1$.
Fig.~\ref{fig:zjets} shows the measured differential inclusive jet 
cross sections as a function of $p_T^{jet}$ in $Z$+jets production for
$N_{jet}\ge1$ and $N_{jet}\ge2$ and the total cross sections,
$\sigma_{N_{jet}}$, for $Z$+jets events up to the $N_{jet}\ge3$ bin.
Good agreement was observed between data and NLO pQCD
predictions from MCFM~\cite{MCFM} up to the $N_{jet}\ge2$ bin where
NLO predictions are available.
The ratio of the data to the LO pQCD calculations indicates that the
LO pQCD predictions underestimate the data by a factor of about 1.4
and this factor is constant over inclusive jet multiplicities up to
the $N_{jet}\ge3$ bin.


CDF has recently updated the measurement on $Z+b$-jet production
using 1.5 fb$^{-1}$ of data.
The measurement was made using
jets with $E_T>20$ GeV and $|\eta|<1.5$ tagged as $b$-jets by the
secondary vertex algorithm in $Z\to e^+e^-$ and $Z\to\mu^{+}\mu^{-}$
events
and the results are summarized in Table.~\ref{tab:zb}.
The measured $Z+b$-jet cross section and its fractions in the $Z$
events and $Z$+jets events are found to be somewhat higher than the
NLO pQCD predictions, and the $Z+b$-jet fractions are in
better agreement with predictions from Pythia Tune A.
The differences between the NLO predictions and Pythia are being
investigated.

\begin{table}
\caption{Results on the $Z+b$-jet production.}
{\small
\begin{tabular}{ccccc}
\hline\hline
 & CDF Data & Pythia & NLO  & NLO+UE       \\
 &          &        &      & $\!\!\!$+Hadronization $\!\!\!\!\!\!$\\
\hline
$\sigma(Z+b\mbox{-jet})$  &  
  $0.94\pm0.15\pm0.15$ pb & 
  n.a.    & 0.51~pb  &  0.56~pb    \\
$\sigma(Z+b\mbox{-jet})/\sigma(Z)$       &  
  $0.369\pm0.057\pm0.055$ \% & 
  0.35~\%  &  0.21~\% &  0.23~\%   \\
  $\!\!\!\!\!\! \sigma(Z+b\mbox{-jet})/\sigma(Z+\mbox{jet}) \!\!\!\!\!\!$ &  
  $2.35\pm0.36\pm0.45$ \%    & 
  2.18~\%  &  1.88~\% &  1.77~\%   \\
\hline\hline
\end{tabular}
}
\label{tab:zb}
\end{table}

\section{Summary}

CDF has a broad program on jet physics including the measurements on
inclusive jet, dijet, $b\bar b$ dijet and boson+($b$-)jets production,
and is making a significant impact on better understanding of jet
production and QCD.
These measurements provide tests of pQCD calculations and Monte Carlo
event generators, and constraints on the proton PDFs. 
QCD processes are often important backgrounds to electroweak and
possible new physics processes, and thus better
understanding of QCD processes will enhance the potential for new
physics discoveries at the Tevatron and also at the upcoming LHC.


\begin{thebibliography}{99}
\bibitem{CDFIncJetKt} A. Abulencia {\it et al.}, Phys. Rev. D 75,
  092006  (2007).
\bibitem{tuneA} R. Field, presented at Fermilab ME/MC Tuning Workshop,
  October 4, 2002.
\bibitem{Jimmy} J.~M.~Butterworth {\it et al.}, 
  Z.\ Phys.\  C {\bf 72}, 637 (1996).
\bibitem{MCNLO} S.~Frixione and B.~R.~Webber,
  JHEP {\bf 0206}, 029 (2002).
\bibitem{MCFM} J.M. Cambell {\it et al.}, Phys. Rev. D {\bf 69},
  074021 (2004).
\bibitem{CKKW} S.~Catani {\it et al.},
  J. High Energy Phys. {\bf 0111}, 063 (2001);
  F.~Krauss,
  J. High Energy Phys. {\bf 0208}, 015 (2002).
\bibitem{Alpgen} M.~L.~Mangano {\it et al.}, 
  JHEP {\bf 0307}, 001 (2003).
\end{thebibliography}
\end{document}